\documentstyle[amssymb,preprint,aps,12pt]{revtex}
\begin{document}
\title{Thermodynamic Density Matrix Renormalization Group Study of the Magnetic
Susceptibility of Half-Integer Quantum Spin Chains}
\author{S. Moukouri and L. G. Caron}
\address{Centre de recherche en physique du solide and d\'epartement de physique,\\
Universit\'e de Sherbrooke, Sherbrooke, Qu\'ebec, Canada J1K 2R1}
\date{23 September 1996}
\maketitle

\begin{abstract}
It is shown that  White's density matrix renormalization group technique can
be adapted to obtain thermodynamic quantities. As an illustration, the
magnetic susceptibility of Heisenberg $S=1/2$ and $S=3/2$ spin chains are
computed. A careful finite size analysis is made to determine the range of
temperatures where the results are reliable. For the $S=1/2$ chain, the
comparison with the exact Bethe ansatz curve shows an agreement within $1\%$
down to $T=0.05J$.

\end{abstract}


One-dimensional quantum spin systems have been extensively studied for many
years. In particular, a number of exact results for thermodynamic quantities
were obtained for the $S=1/2$ \cite{takahashi}and $S=\infty $\cite{fisher}
systems. Unfortunately, the intermediate spin systems still offer resistance
to analytical methods. The solution of the classical spin model is therefore
often used to fit experimental curves. While the magnetic susceptibilty $%
\chi $ of the $S=5/2$ tetra-methyl manganese chloride (TMMC) was well
understood using this model\cite{hutchings}, that of $CsVCl_3$, which is a
physical realization of a one-dimensional Heisenberg $S=3/2$ chain, has
shown significant discrepancy from it\cite{niel}\cite{itoh}. 
Althought quantum effects
in a $S=3/2$ system are expected to be weaker than in a $S=1/2$, they are
believed to be non negligible.

The purpose of this paper is to show that the density matrix renormalization
group method (DMRG)\cite{white1} can be applied reliably for systems at
finite temperatures, the only limitation being of numerical origin at the
lower temperatures. Then, after a test on the $S=1/2$ Heisenberg chain, we
will compute $\chi $ for the $S=3/2$ chain. 

Apart from the Wilson solution
of the Kondo problem\cite{wilson}, earlier attempts to implement numerical
renormalization group techniques to systems at finite temperature has led to
unsatisfactorily results. Chui and Bray\cite{chui} had little success on
their RG study of the one-dimensional Hubbard model. A test by Hirsch\cite
{hirsch} on the simpler $S=1/2$ Heisenberg chain led also to poor results.
The advent of the DMRG has provided a framework offering new expectations.
The DMRG has proven its remarkable efficiency in the calculation of the
ground-state properties of many one-dimensional quantum hamiltonians. White%
\cite{white1} showed formally that the method can easily be generalized to
systems at finite temperatures. One only has to target several excited
states when building the reduced density matrix. However, a straightforward
application of this idea presents some difficulties. For a fixed number $m$
of states in the two largest blocks, when the number of the target states $M$
is increased, the truncation error increases. To retain good accuracy, $m$
must be at least greater than $M$. For the $S=1/2$ Heisenberg spin chain the
value of $M$ necessary to obtain the thermodynamics quantities for $T$ (in
units of the exchange constant) in the range $0-0.5$ is roughly estimated to
be several thousands for lattices of less than $N=30$ sites. Although this
value is small compared to the number of states of the full Hilbert space,
the computation of the $M$ states through the standard block-Lancz\"os or
block-Davidson algorithms is impractical because these require a very large
amount of memory as well as CPU time. These difficulties can be circumvented
(non rigorously) if the $M$ states are used not for the calculation of
thermodynamic quantities but only to generate approximate hamiltonian
matrices whose sizes are kept within manageable values so as to allow
complete diagonalization yet large enough for accurate thermodynamic
calculations as discussed above. In such a situation, we found that a few
tens of target states can be used to construct reliable hamiltonian matrices
that can be diagonalized using dense matrix algorithms.

We will illustrate these ideas on the one-dimensional Heisenberg model: 
\begin{equation}
H=J_S\sum_i{\bf S}_i^{}\cdot {\bf S}_{i+1}^{}  \eqnum{1}
\end{equation}
We will first study the case $S=1/2$, $J_S=J=1$. A long time ago, Bonner and
Fisher\cite{bonner} computed the magnetic susceptibility of chain lengths of
up to 11 sites by exact diagonalization. They used periodic boundary
conditions (PBC) which they found to converge much more rapidly to the
infinite chain limit. Since then, the Bonner-Fisher curve is used by
experimentalists to estimate the exchange constant of physical systems. But
recently, it was noticed\cite{eggert} that the results of Bonner and Fisher,
while excellent at high temperatures, turn out to deviate from the Bethe
ansatz curve at lower temperatures (below $T\approx 0.2$). This may be
related to the correlation length which increases at low temperatures. As a
consequence, one needs to diagonalize longer chains. We learn from this that
the method of Bonner and Fisher would presumably lead to even worse results
for systems with more than two degrees of freedom per site.

In the DMRG, if $B$ is the block whose size is increased by one at each step
and $\bullet $ represents an added site, then, using PBC, the superblock is $%
B\bullet B\bullet $. Our procedure goes as follows: $(i)$ We start by
obtaining all the eigenvalues (through a dense matrix algorithm) and
thermodynamic averages of $BB$ with $N$ sites, 
which, in the first iteration, is taken to
be the largest possible ($N=14$ in the case of $S=1/2$ chain on our
computer). Note that this step which is necessary for the computation of
thermodynamic quantities is absent in the $T=0$ DMRG algorithm. This choice
of $BB$ for the measurements instead of the superblock is motivated by 
the manageable size of the matrices.
 As usual, the states of $BB$ as well as those of the superblock
are labelled by the z-component of the total spin and the use of parity
symmetry allows us to compute only the subspaces $S_T^z=0,1,...N/2.$ $(ii)$
We next build the superblock which has $N+2$ sites. 
Its $M$ lowest states are computed by the
block-Davidson algorithm. For a fixed $m$, $M$ is chosen to be the largest
value for which the truncation error $p(m)$ is less than a desired accuracy,
say $5\times 10^{-4}.$ $(iii)$ We form and diagonalize the reduced density
matrix of $B\bullet $ built from these $M$ states. We make the
transformation $B=B\bullet $, and return to $(i)$. It is important to point
out that our optimization procedure in part $(ii)$ implies, reciprocally,
that it is only the $M$ targeted states that are calculated with high
accuracy. It is, however, quite clear that even if only one target state is
used, say the ground state of the system, the internal consistency of the
DMRG is such that many of the excited states of $BB$ are also obtained with
reasonable accuracy. Therefore, if many states are targeted, one expects
that the excited states of $BB$ above the $M$ targeted states will be
reasonably well estimated. The accuracy will, of course, certainly decrease
as one goes higher in excited energy. However, the weight of these states in
the partition function also decreases, thus compensating for the loss of
accuracy. There is a simple test to this statement. For 
 $m=128$, we compared the energy of the states $M+1$, $M+2$ 
...just above the
cut-off $M$ (for $M=20$) to their energies when $M$ is increased enough to
include them as target states. For the case $N=16$, the difference are in the
order of $10^{-3}$ for the $5$ lowest states above the cut-off. But as $M$ is
further increased, the truncation error gets larger. Thus, the accuracy is
reduced even on the target states if $M$ is to larger for a fixed $m$.

Let $n$ and $\Delta E_n=E_n-E_G$ be the number and the energy range of the
retained eigenvalues of $BB$ respectively. $E_G$ is the ground state energy
and $E_n$ is the highest eigenvalue kept of the subspace $S_T^z=0$. To
obtain good accuracy, the contribution to the partition function of the
discarded eigenstates must be much smaller than that of the $n$ states.
After some steps of the DMRG, since we keep only a small fraction of states,
the number of discarded states is approximately equal to the dimension of
the full Hilbert space $2^N$. Hence, at a fixed temperature, despite the
exponentially vanishing contribution of each rejected state, their total
contribution to the partition function is not necessarily negligible because
of their exponentially large number. An upper bound to this contribution is
roughly $2^N\exp (-\beta \Delta E_n)$. In practice we did not compute all
the subspaces $S_T^z=0,1,...N/2$; we have found that the energies of the
lowest states of the subspaces with $S_T^z\geq 7$ are higher than $E_n$ for
our chosen values of $m$ and $M$ and so these subspaces were not retained.
The total number of states involved 
in $BB$ for the measurements at each DMRG step is around $n=12000$.
Despite the truncation of states in the reduced density matrix, these
truncated states would contribute mainly to form eigenvalues higher than $E_n
$ in $BB$. In the present the calculations, we took, $m=128$ and $M=30$. The
target states are distributed between the different subspaces as follows: $10
$ states for the $S_T^z=0$ subspace, $8$ for $S_T^z=1$, $4$ for $S_T^z=2$,
and $2$ in each $S_T^z=3$ to $6$. This choice, which seems somewhat
arbitrary, is motivated by the fact that the subspaces having the lower $%
\left| S_T^z\right| $ contain the larger number of low lying states. The
largest matrix size we diagonalized was $3432$. The most demanding part of
the computation obviously involved the dense matrix 
diagonalization of $BB$. This part of algorithm is about $50$ times 
longer than the calculation of the $M$ states of the superblock using the
Davidson algorithm.

Fig. 1 shows the results for $\chi $ in chain of increasing even lengths
from $N=4$ up to $N=30$. When $N\leq 14$, the DMRG is equivalent to exact
diagonalization since all the states are kept. The truncation starts at $%
N=16 $ and, in this situation, the range of temperatures where we can
calculate $\chi $ with almost no finite size effects is $[\Delta
E_N,T_{max}] $, where $\Delta E_N$ is the spin gap observed at that chain
length and $T_{max}$, which is the consequence of the elimination of high
lying states, is defined such that $2^N\exp (-\beta _{\max }\Delta
E_n)\approx 10^{-3}$. The absence of finite size effects (less than $0.1\%$
when $N\geq 14$) above $\Delta E_N$ is due to the application of PBC and to
the short range nature of the interaction. In fact for a given size $N$,
when $T$ is greater than $\Delta E_N$, the value of the susceptibility is
very close to that of the infinite chain. Finite size effects start when $T$
is in the vicinity or lower than $\Delta E_N$. For the open boundary
conditions (OBC), these would be much more significant; one would need in 
principle to
extrapolate the results to infinity. The main advantage of the OBC is that $%
p(m)$ would be smaller. As we iterate the DMRG, $T_{max}$ and the width $%
T_{max}-\Delta E_N$ get smaller because of the renormalization. A natural
criterion when to stop the DMRG iterations appears be $T_{max}
{< \atop \sim}
\Delta E_N$. But, it should be noticed that the values of $\chi $ for $T
{< \atop \sim}
 \Delta E_N$, where finite size effects are significant, are also
accessible by extrapolation; the best fit is obtained with rational
functions. We were able to calculate $\chi $ down to $T\approx 0.05$. Below
this temperature, changes in the values of $\chi $ at each $N$ are of the
order of $p(m)$ so that our extrapolation became unreliable. The agreement
with the exact Bethe Ansatz result is quite excellent as shown in Fig 2. The
DMRG curve confirms the inflection point found in ref\cite{eggert} at $%
T=0.087$. The maximum deviation from the exact value is found to be less
than $1\%$. The algorithm presented above is an infinite lattice one; the
accuracy could certainly be increased by applying the finite system method%
\cite{white1}.

The same analysis is now applied to the $S=3/2$ chain. In this case, there
are four degrees of freedom per site. The longest chain that we were able to
diagonalize exactly is $N=7$. For the DMRG calculations, we took $m=150$, $%
p(m)\approx 7\times $ $10^{-4}$, $M=20$ and the maximum chain length $N=16$.
We have taken $10$ subspaces corresponding for the values of $S_T^z$ ranging
from $0$ to $9$. We retained about $n=20000$ states at each step. In order
to allow actual comparison with the $S=1/2$ and $S=\infty $ chains, the
exchange constant and the $g$ factor were chosen as follows: $S^2J_S=J/4$
and $Sg_S=g/2$. The results for $\chi $ are shown in Fig 3. The limiting
susceptibility is bracketed by the curves of odd and even $N$ obtained by
exact diagonalization. We could not expect to get the same accuracy as in
the $S=1/2$ case since the truncation of the Hilbert space started after $N=6
$ only. The results for the high temperatures are not completely free of
size effects. These are reduced by taking the average of $N=6$ and $7$. In
conformity with the gapless spectrum, $\chi $ seems to extrapolate to a
finite value at $T=0$. We found $\chi =0.12$ at $T=0.04$. But we have not
made an extrapolation to $T=0$; an eventual infinite slope, as for the $S=1/2
$ chain\cite{eggert}, could have been missed. We note that our results are
in agreement with a recent Monte Carlo study\cite{sandvik}. 
The Monte Carlo calculation has reached a lower
temperature than the present DMRG study. But, we believe
that the results presented here can significantly be improved by 
using a powerful computer rather than our modest workstation. In Fig 4, the
calculated $\chi $ of the $S=3/2$ chain is compared with the Bethe ansatz
curve of the $S=1/2$ and Fisher's curve of the $S=\infty $ chain. As
expected, the Curie-Weiss behavior starts at lower temperature than in the $%
S=1/2$ chain but at higher temperature than the classical spin.

In conclusion, we have shown that the DMRG is able to produce a static
susceptibility of quantum spin chains which compares favorably with the
available exact result in an extensive range of temperatures. The
calculation of other thermodynamic quantities is straighforward; care must
be taken to watch for finite size effects which depend on the quantity
calculated. The method can presumably be used in the investigation of
frustrated spin systems as well as for
fermion systems such as the Hubbard or $t-J$ models; it will probably not
have the difficulty that hampers the Monte Carlo method
. There is a foreseeable problem  when PBC are used
associated with accidental degeneracies. 

The calculations were made
on a workstation having $128 Mb$ memory and a speed of $25 Mflops$. Each DMRG 
step
required approximatively $1200$ minutes.

We benefited from earlier discussions with Liang Chen. We acknowledge
helpful correspondence with S. Eggert. We wish to thank A-M. S. Tremblay for
many useful discussions. This work was supported by a grant from the Natural
Sciences and Engineering research Council (NSERC) of Canada and the Fonds
pour la formation de Chercheurs et d'Aide \`a la Recherche (FCAR) of the
Qu\'ebec government.

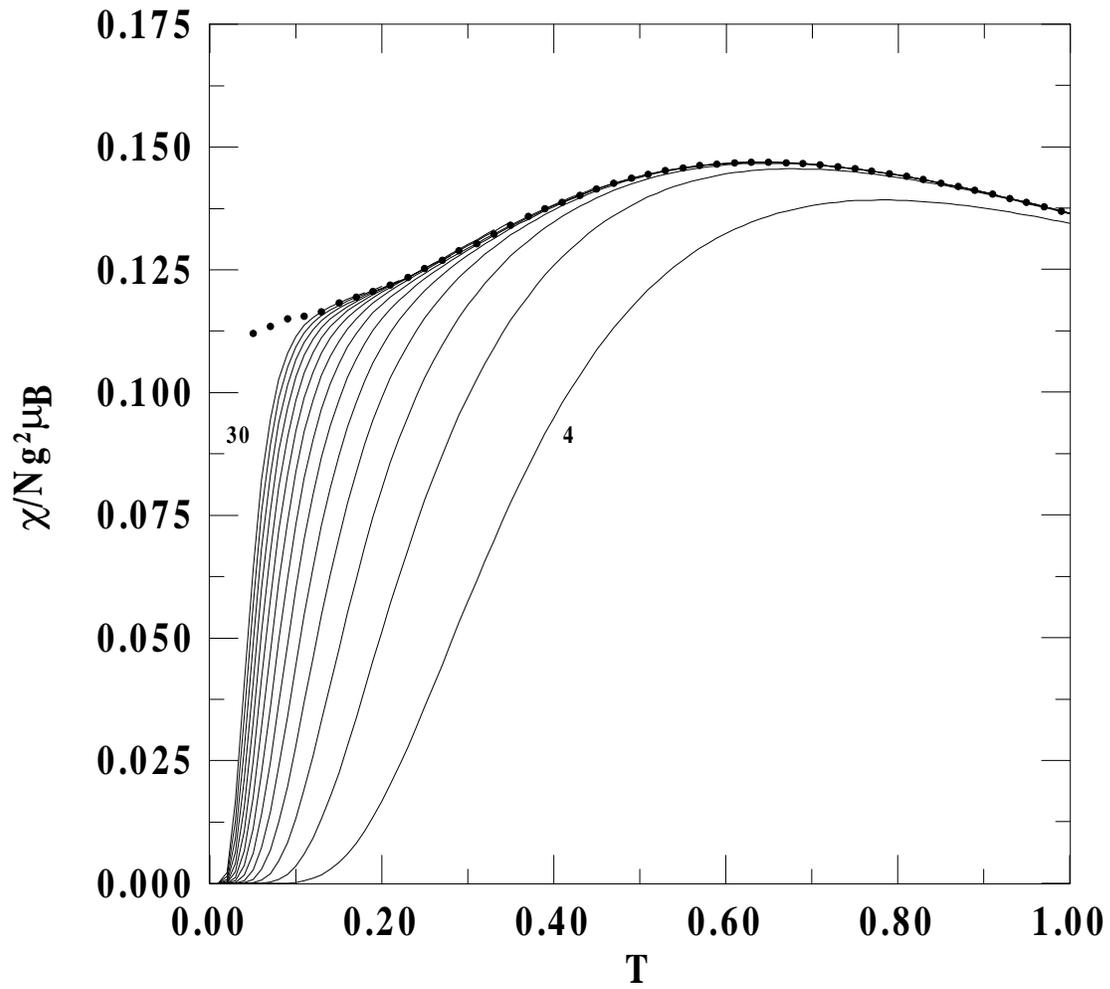
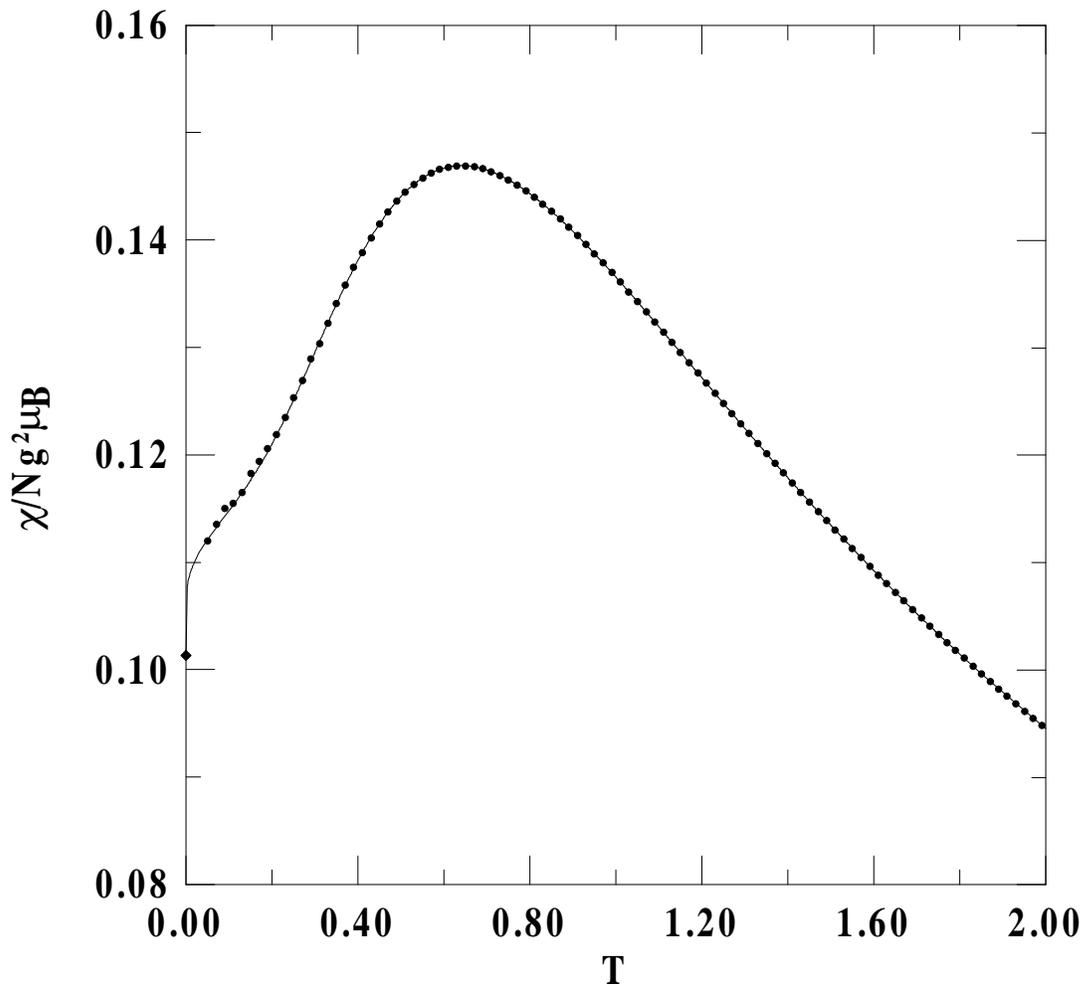
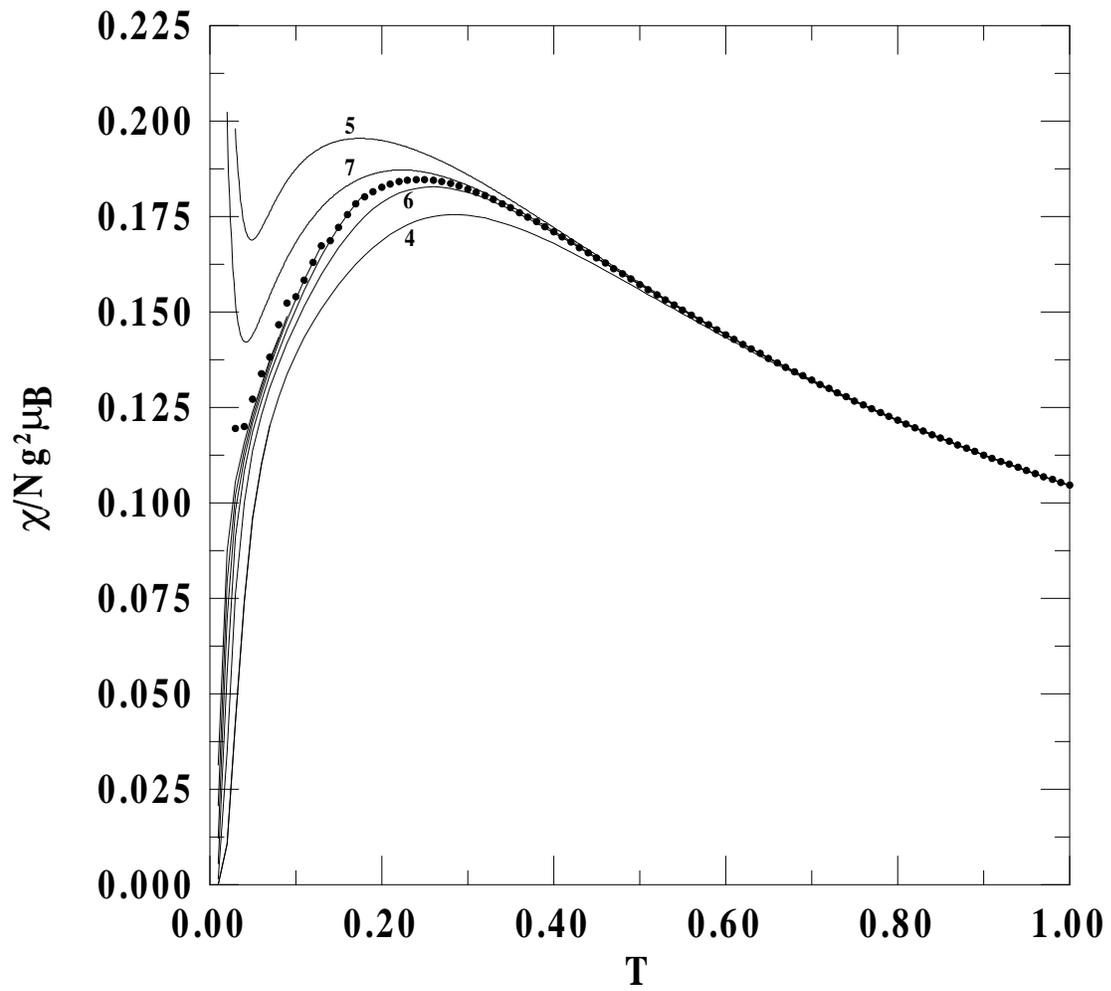
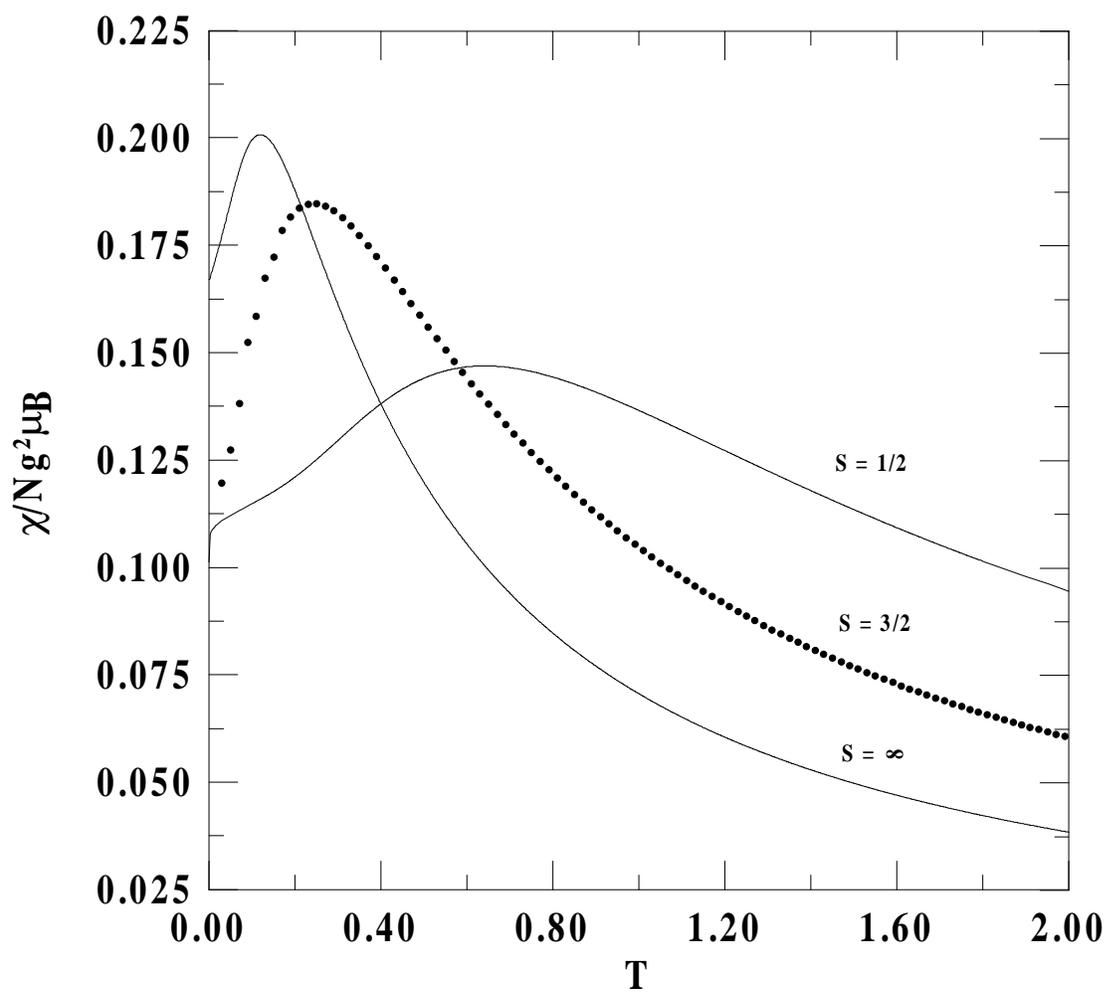
\end{document}